# A Reference Model for Collaborative Business Intelligence Virtual Assistants

Olga Cherednichenko [1], Fahad Muhammad [1], Jérôme Darmont [1] and Cécile Favre [1]

[1] *Univ Lyon, Univ_Lyon 2, UR ERIC – 5 avenue Mendès France, 69676 Bron Cedex, France*

**Abstract**
Collaborative Business Analysis (CBA) is a methodology that involves bringing together different stakeholders, including business users, analysts, and technical specialists, to collaboratively analyze data and gain insights into business operations. The primary objective of CBA is to encourage knowledge sharing and collaboration between the different groups involved in business analysis, as this can lead to a more comprehensive understanding of the data and better decision-making. CBA typically involves a range of activities, including data gathering and analysis, brainstorming, problem-solving, decision-making and knowledge sharing. These activities may take place through various channels, such as in-person meetings, virtual collaboration tools or online forums. This paper deals with virtual collaboration tools as an important part of Business Intelligence (BI) platform. Collaborative Business Intelligence (CBI) tools are becoming more user-friendly, accessible, and flexible, allowing users to customize their experience and adapt to their specific needs. The goal of a virtual assistant is to make data exploration more accessible to a wider range of users and to reduce the time and effort required for data analysis. It describes the unified business intelligence semantic model, coupled with a data warehouse and collaborative unit to employ data mining technology. Moreover, we propose a virtual assistant for CBI and a reference model of virtual tools for CBI, which consists of three components: conversational, data exploration and recommendation agents. We believe that the allocation of these three functional tasks allows you to structure the CBI issue and apply relevant and productive models for human-like dialogue, text-to-command transferring, and recommendations simultaneously. The complex approach based on these three points gives the basis for virtual tool for collaboration. CBI encourages people, processes, and technology to enable everyone sharing and leveraging collective expertise, knowledge and data to gain valuable insights for making better decisions. This allows to respond more quickly and effectively to changes in the market or internal operations and improve the progress.

## 1. Introduction

Data analysis help us to make better decisions based on the insights and patterns we discover in data. By analyzing data, we can identify trends, relationships and patterns that would be difficult or impossible to discern otherwise. By using data to inform our decisions, we can improve outcomes and avoid costly mistakes. In addition, data analysis is increasingly important in today's world of big data, where there is an abundance of information available, but it can be overwhelming to make sense of it all. Data analysis tools and techniques help process and understand this vast amount of information,

---

making it more manageable and actionable. Collaborative Business Analysis (CBA) is a process of analyzing data and making business decisions collaboratively. According to the analysis, the need for data research is not limited to businesses seeking profit or governments tackling national issues, but also includes individuals and society as a whole. These groups require data analysis to make socially significant or personal decisions. However, it can be challenging to facilitate collaboration between potential users, decision-makers, and technical specialists in data analysis. The BI4people project [1] aims to address these issues by leveraging Online Analytical Processing (OLAP) to provide interactive analysis and data visualization through a software-as-a-service model. This allows a broader audience to access data warehousing, including multi-source, heterogeneous data integration.

Collaborative Business Intelligence (CBI) involves using social networks, quizzes, brainstorming sessions, and even simple chats between two or three coworkers on teams to collectively solve problems [2]. The key concept of CBI is to bring people together in a virtual or physical space and encourage them to share their opinions and comments for the common good. Additionally, the reuse of other collaborators' comments or results is also considered part of CBI, which ultimately leads to a more comprehensive approach to Business Intelligence (BI).

CBI and Collective Intelligence (CI) are related concepts that both involve bringing people together to work collaboratively and share knowledge. However, there are some important differences between the two. CI is a concept that refers to the ability of a group or network to collectively solve problems, make decisions and generate insights that are often beyond the abilities of individuals working alone. The idea is that, by tapping into the collective knowledge and expertise of a group, we can achieve better solutions and make better decisions than if working in isolation. CBI specifically refers to the use of collaborative tools and methods to support BI processes. This may include social networks, brainstorming sessions, chat applications and other technologies that facilitate knowledge sharing and collaboration between business users, analysts and technical specialists. The goal of CBI is to leverage the collective knowledge and insights of the group to improve decision-making, drive business outcomes and gain a competitive advantage.

As data exploration is the core of BI, it is important to design powerful and efficient tools to support BI processes, such as data analysis or visualization. From the other side, it can be extremely hard to manage data analysis for inexperienced user. Analysis shows that some solutions from Artificial Intelligence (AI) can help with an issue. Machine Learning (ML) and Natural Language Processing (NLP) can help organizations to automate and optimize their data analysis processes and make more accurate and informed decisions based on the insights derived from data.

A chatbot is a computer program that can simulate human conversation through text or voice interactions with users [3]. Chatbots are typically designed to perform specific tasks or provide information to users. They are usually rule-based, meaning they follow pre-defined scripts or decision trees to respond to user inputs [4,5]. There are some chatbots and Conversational Agents (CAs) that allow users to combine commands through nested conversations to accomplish open-ended data analysis tasks.

1. IBM Watson Assistant [6] allows users to create complex conversational flows and combine various commands to achieve data science tasks.

2. The Dialogflow platform [7] provides a natural language understanding system that enables users to create conversation flows and nested dialogues to handle complex data science tasks.

3. Rasa [8] is an open-source conversational AI framework that enables developers to build AI assistants with advanced NLP capabilities and flexible dialogue management.

4. The Botpress platform [9] provides a workflow builder that enables users to design chatbots with nested conversations, custom scripts and integrations with third-party services.

5. The Wit.ai platform [10] offers an NLP system that allows developers to create intelligent chatbots, voice assistants and other conversational AI applications with complex dialogue flows.

These are just to name a few examples of chatbots and CAs that allow users to combine commands through nested conversations to accomplish open-ended data science tasks.

Following by the cutting-edge tendency the given paper focus on how the conversational chatbots can improve the CBI tools and bring new opportunities for businesses, governments, and citizens. We suggest a reference model of virtual collaborative assistant for CBI, which consists of three components: conversational, data exploration and recommendation agents. We believe that the allocation of these three functional tasks allows you to structure the CBI issue and apply relevant and productive models for human-like dialogue, text-to-command transferring, and recommendations simultaneously. The complex approach based on these three points gives the basis for virtual tool for collaboration.

The rest of the paper is structured as follows. The next section depicts the concept of CBI and reasons to use chatbots for it. The research questions are highlighted. Then the state-of-the-art is presented. The fourth section describes a brief summary of the methods. And the fifth section represents the CBI virtual assistant reference model. The next section provides a discussion and, finally, we conclude our results.

## 2. Background

### 2.1. The CBI concept

CBA is an effective way to optimize business analysis processes, enhance collaboration and drive better business outcomes. CBA usually involves the use of modern collaborative tools, such as social networks, brainstorming sessions, and chat applications, to enable stakeholders to work together in real-time, regardless of their physical location. By using these tools, stakeholders can share their thoughts, ideas, and insights in a more interactive and immediate way, leading to a more efficient and effective analysis process. CBI involves enabling users to share data, insights, and perspectives with one another in order to improve decision-making and drive better business outcomes [2, 11, 12]. The main points of the concept of CBI can be summarized as follows.
- Collaboration: CBI is centered around the idea of collaboration and knowledge sharing among team members. By facilitating communication and collaboration, organizations can create a culture of data-driven decision-making.
- Data sharing: CBI involves sharing data across teams and departments. This means breaking down data silos and ensuring that everyone has access to the same data, regardless of where it is stored.
- Visualization: CBI leverages data visualization tools to make data more accessible and easier to understand. This allows team members to quickly identify trends and patterns and make informed decisions.
- Agility: CBI is agile, meaning that it enables teams to quickly adapt to changing business needs and make decisions based on the latest data.
- Continuous Improvement: CBI is a continuous process of improvement, with team members working together to identify opportunities for improvement and make changes to their processes accordingly.

Thus, CBI is a powerful approach to data-driven decision-making that enables organizations to harness the collective knowledge and expertise of their teams in order to drive better business outcomes [12, 13].

### 2.2. The chatbots

Chatbots are computer programs that use AI to simulate conversation with human users [14]. There are several reasons why chatbots are becoming increasingly popular for collaborative tools.
- Convenience: Chatbots provide a convenient way for users to access information and perform tasks without having to leave the communication platform. This makes it easy for team members to collaborate and share information without having to switch between multiple tools.
- Efficiency: Chatbots can automate routine tasks and processes, freeing up time for team members to focus on more high-value activities. This can improve productivity and help teams work more efficiently.
- Personalization: Chatbots can be customized to meet the specific needs of individual users, providing personalized experiences that can improve engagement and adoption.
- Scalability: Chatbots can handle large volumes of requests and interactions simultaneously, making them well-suited for collaborative tools that serve a large number of users.

- Availability: Chatbots are available 24/7, providing users with access to information and support around the clock. This can be especially valuable for global teams that work across different time zones.

Chatbots provide a convenient and efficient ways for users to access information and perform tasks within collaborative tools. By automating routine tasks and providing personalized experiences, chatbots can help teams work more effectively and achieve their goals more efficiently.

In this research we focus on the development of a CBI framework in terms of conversational assistant concept (i.e., a chatbot). The main goal of our research is to investigate and improve the CBI tools based on virtual assistants. Towards this aim, we make out the following research questions.

1. What is the main scenery for a collaborative tool on a virtual platform?
2. What type of CA is the most relevant as a CBI assistance tool?
3. What AI models and natural language understanding should feature in a CBI virtual assistant?

## 3. Related Works

There exist multiple approaches to implementing CBI [11]. There are Social BI, Mobile BI, Cloud-based BI, Self-service BI, and Embedded BI [12].

Social BI leverages social media platforms to gather insights and share information with a wider audience [15]. This approach can be used to facilitate collaboration and information-sharing across teams and departments. It involves gathering and analyzing data from social media channels, such as Twitter, Facebook, LinkedIn, and others, to gain insights into customer behavior, market trends and brand reputation. Social BI also involves collaboration among team members, such as sharing and discussing data, visualizations, and reports to make better decisions.

Mobile BI (MBI) [16] provides access to BI tools and information through mobile devices such as smartphones and tablets. This approach can be useful for remote teams or teams that are frequently on-the-go.

With cloud-based BI (CBBI) [17], data are stored in a centralized location in the cloud, rather than on local servers or hard drives. This makes it easier for organizations to access and share data, regardless of their location or device. CBBI tools are typically accessible through Web browsers or mobile applications, which provide users with a simple and convenient way to access data and perform analysis. CBBI can offer several benefits, such as cost savings, scalability, and flexibility. Organizations can avoid the high costs of building and maintaining an on-premises BI infrastructure and can scale BI capabilities up or down as needed. Additionally, cloud-based BI allows users to work from anywhere with an Internet connection, enabling remote teams to collaborate and share insights more easily.

Self-service BI (SSBI) refers to the approach of empowering business users to create their own reports, dashboards, and analyses, without relying on IT or data analysts [18]. It enables users to access and analyze data without the need for technical skills or assistance from IT staff. SSBI tools usually have user-friendly interfaces that allow users to access and manipulate data with ease, including drag-and-drop features, visualization tools and NLP capabilities. This approach allows business users to answer their own questions, explore data and generate insights on their own, which can help in faster decision-making and better business outcomes.

Eventually, embedded BI (EBI) integrates BI tools and insights directly into other business applications and workflows [19]. This can help improve collaboration by providing users with easy access to relevant data and insights within the context of their work. The main advantage of EBI is that it allows users to access analytics and insights directly within the applications they already use, without having to switch between different tools or interfaces. This can improve productivity, as users can make decisions more quickly and easily, and can also increase adoption of BI capabilities by making them more accessible.

Moreover, chatbots can be powerful tools for facilitating CBI, by providing users with quick and easy access to data and insights, improving collaboration, feedback and helping to democratize data and insights across an organization. For example, in a research article [20], the authors suggest that chatbots can be used to facilitate collaboration and knowledge sharing among team members, particularly for tasks such as information retrieval and decision-making. Other experts in the field have also advocated for the use of chatbots in CBI as a way to improve collaboration, increase accessibility and streamline workflows [21, 22].

Chatbots can fall into different categories depending on their functionality and the type of collaboration they support [4, 14, 22]. Chatbots can be used to provide insights, recommendations or predictions based on the data available. Collaborative chatbots can be used to assign tasks, schedule meetings, or share information among team members. For example, a chatbot could be used to gather feedback from users on a dashboard or report, or to facilitate discussions between team members about key insights or metrics. Chatbots can be used to send notifications and alerts to users based on predefined triggers, such as changes in data or anomalies in key metrics. This can help teams to stay informed and make timely decisions based on real-time information.

A chatbot is a type of CA, but not all CAs are chatbots. CA is a broader term that encompasses any type of computer program or system that can engage in natural language interactions with users [21, 23]. CAs can be rule-based or rely on ML and NLP techniques to understand and respond to user inputs.

The Iris CA is a dialogue system developed by Stanford University's NLP Group [24]. It is designed to carry out text-based conversations with users in natural language. Iris is designed to help data scientists and other users interact with data and ML models through natural language conversations [24]. Iris can be integrated into various platforms and tools such as Slack, Jupiter Notebooks and APIs. Users can ask Iris questions about data, such as "What is the average salary for employees in my company?" or "What is the distribution of ages in my dataset?" Iris can also assist users in creating visualizations and analyzing data using ML models. Iris uses NLP and ML techniques to understand the user's intent and generate relevant responses. The system is designed to learn from user interactions and improve over time. Iris can also handle complex queries and handle multiple data sources.

Some similar CAs are Apple's Siri [25], Amazon's Alexa [26] and Google Assistant [27]. These virtual assistants also use NLP to help users completing tasks, answer questions and perform actions. There are also several chatbot platforms available, such as Dialogflow [28] and Microsoft Bot Framework [29], which allow developers to create their own CAs.

Furthermore, there are several neural network architectures that are commonly used for designing CAs. The most spread are Recurrent Neural Networks (RNNs), Long Short-Term Memory (LSTM), Generative Adversarial Networks (GANs), the seq2seq models and transformers [4, 24, 30, 31]. RNNs are a type of neural network that is well-suited for processing sequential data, such as text [32]. They are often used for language modeling, which involves predicting the next word in a sequence given the preceding words. In the context of CAs, RNNs can be used to generate responses to user inputs. LSTM networks are a type of RNNs that are designed to handle the problem of vanishing gradients, which can occur when training deep neural networks [33]. LSTMs are particularly effective at modeling long-term dependencies in sequential data, which makes them well-suited for CAs that need to understand the context of a conversation. GANs are a type of neural network that can be used to generate realistic synthetic data, such as text [34]. In the context of CAs, GANs can be used to generate responses to user inputs that are more diverse and creative than those generated by traditional rule-based or template-based approaches.

The seq2seq model is a neural network architecture commonly used for NLP tasks such as machine translation and text summarization [24, 35]. For example, in Iris [24], the seq2seq model is used to process user queries and generate responses. The attention mechanism helps the model to focus on relevant parts of the input sequence when generating the output sequence. In addition to the seq2seq model, Iris also uses Reinforcement Learning (RL) to improve its performance over time [24]. RL is a type of ML where an agent learns to take actions in an environment in order to maximize a cumulative reward signal [35]. In traditional RL, the reward signal is provided by the environment, but in RL with human feedback, the reward signal is provided by a human expert. Thus, RL is a commonly used approach to build CAs. In Iris, RL is used to improve the accuracy and relevance of the responses generated by the seq2seq model. Finally, transformers are a type of neural network architecture that is designed to process sequences of input data in parallel, rather than sequentially like RNNs [36]. This makes them much faster and more efficient than RNNs, which makes them well-suited for large-scale CAs.

These are just a few examples of the neural network architectures that are used for CAs. The specific choice of architecture depends on the particular task and dataset at hand, as well as the resources available for training and deployment.

Overall, the field of CA development is rapidly evolving, with a variety of approaches being used to create effective conversational interfaces. One of the most prominent trends is the use of neural networks and deep learning techniques to improve the accuracy and naturalness of CAs. This includes the use of pretrained language models to generate more realistic and human-like responses. In addition, there is a growing interest in the use of CAs for various applications, including customer service, healthcare, education, and entertainment. This has led to the development of specialized CAs for specific domains.

We can conclude that the use of chatbots and CAs is becoming increasingly prevalent in CBI, as they provide a user-friendly interface for retrieving and sharing data. The current trends in CBI include the adoption of cloud-based solutions, mobile BI and the integration of advanced technologies such as NLP and ML. Self-service BI, which empowers users to create their own reports and analyses, is also gaining popularity.

## 4. Methods and Materials

There are various techniques that can be useful for determining the key questions to ask during the exploration process of defining a business problem. Here are some suitable techniques.

1. SWOT analysis: This is a structured framework that helps in identifying the strengths, weaknesses, opportunities, and threats associated with a business problem [37]. It helps identify the important questions related to the business problem.

2. Root cause analysis: This technique identifies the underlying root causes of a problem [38]. It helps identify the key questions related to the root causes that need to be explored during the exploration process.

3. Fishbone diagram: This is a graphical tool that helps visualize the possible causes of a problem [39]. It can be used to identify the key questions related to the causes that need to be explored.

4. Five Whys: This is a simple questioning technique that involves asking "why" five times to get to the root cause of a problem [40]. It helps identify the key questions related to the root cause that needs to be explored.

5. Mind map: This is a visual technique that helps to brainstorm ideas and concepts related to the business problem [41]. It can be used to identify the key questions related to the business problem.

Data exploration is an important component of the BI process, which involves collecting, identifying, and analyzing data to discover meaningful insights and patterns. The main goal of data exploration is to identify key business opportunities and challenges that can drive decision-making and improve business performance. The main steps of data exploration for BI can be summarized as follows.

1. Defining the business problem involves identifying the business area where data exploration can be useful, defining the problem to be solved and determining the key questions to ask during the exploration process.

2. Collecting data involves gathering raw data from various sources, such as database systems, data warehouses, spreadsheets, Web pages, social media, or other relevant sources.

3. Preparing data involves cleaning and formatting data to ensure its accuracy, completeness, consistency, and quality. This step also involves transforming data into a more usable format, such as tables, graphs, charts, and other visualizations.

4. Exploring data involves analyzing data to identify patterns, trends, correlations, and relationships among different variables. This step may involve exploratory analysis techniques, such as clustering, regression analysis, correlation analysis, time-series analysis, and other statistical methods.

5. Communicating insights involves presenting the findings in a clear and concise manner, using visualizations, dashboards, reports, or other forms of communication that can be easily understood by stakeholders.

6. Action planning involves using the insights gained from data exploration to drive action and make informed decisions that can drive business performance. This step may involve developing strategies, setting goals and objectives, and implementing action plans based on the insights gained from data exploration.

Moreover, there are several architectures that are commonly used for CA development.

1. Rule-based architecture: CAs use a set of predefined rules and decision trees to understand user inputs and generate responses [42]. This approach is simple and easy to implement but may not be able to handle complex interactions nor adapt to changes in user behaviour.

2. Finite State Machines (FSM) architecture: CAs use a set of states and transitions to model the conversation flow [43]. This approach is more flexible than rule-based architectures and can handle more complex interactions but may require more development effort to design and maintain the state machine.

3. Natural Language Understanding (NLU) architecture: CAs use ML algorithms to understand user inputs and generate responses [44]. This approach is more advanced and can adapt to changes in user behaviour over time but requires a large amount of training data and computational resources.

4. Hybrid architecture: Many CAs use a combination of the above architectures to balance simplicity, flexibility, and accuracy [3, 31, 44, 45]. For example, a CA may use rule-based techniques for simple interactions and NLU for more complex ones.

Overall, the choice of architecture depends on the specific requirements of the CA and the resources available for development and training.

Fig. 1 shows the three modules that make up a typical task-oriented dialogue agent. These modules are: (1) NLU, which identifies user intents and extracts associated information; (2) Dialogue Management, which tracks the dialogue state to capture all essential information in the conversation and selects the next action based on the current state; (3) Natural Language Generation (NLG), which converts agent actions to natural language responses. In recent years, there has been a growing trend towards creating fully data-driven systems by combining these modules using a deep neural network.

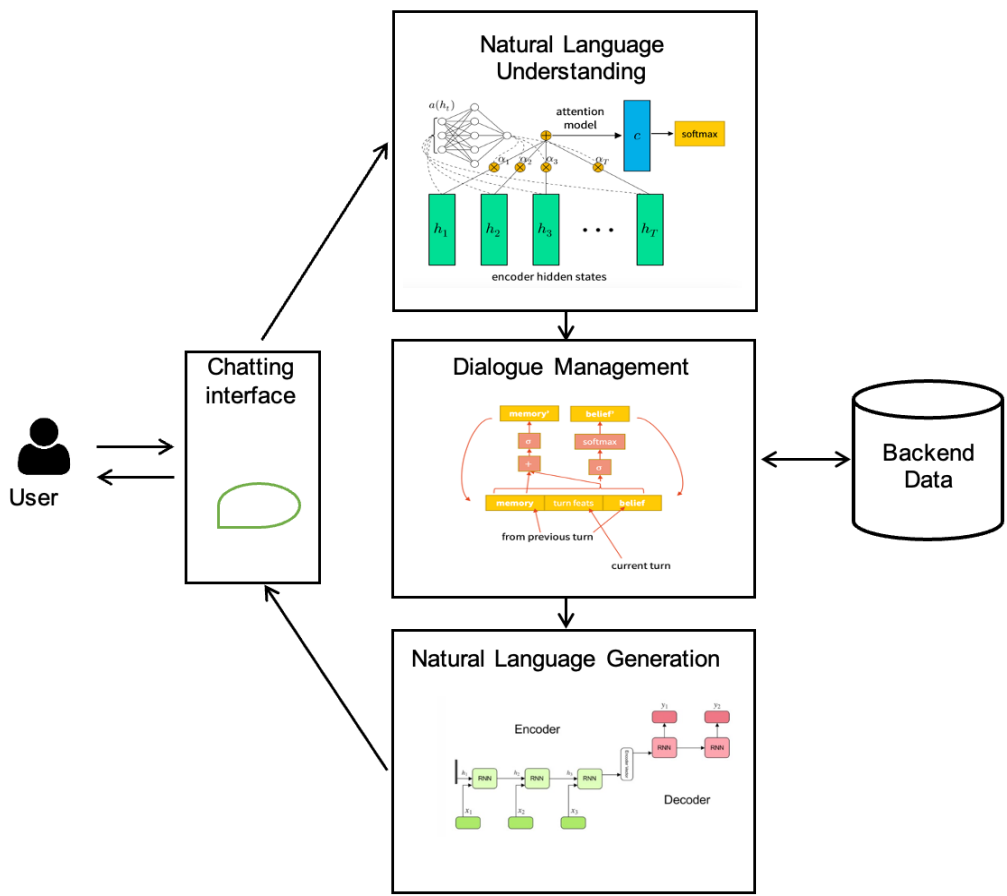

**Figure 1**: Typical dialogue agent

Thus, to choose the appropriate reference model for a CBI virtual assistant, we need to precise the collaborative model, the steps of data exploration and a basic neural network architecture.

## 5. Results

Let us consider in more details the task's features of organizing joint data analysis. It is necessary to consider three main aspects for the implementation of the interactive agent for CBI (Fig. 2). First, the CA must communicate in natural language, not use highly specialized vocabulary and carry on a human-like conversation. Similar tasks are solved by using a neural network architecture such as a Transformer, e.g., ChatGPT. Second, the virtual assistant must be able to convert user instructions into data processing commands (filtering, median calculation, feature grouping, etc.). An example of successful solution to such an issue is the Iris dialogue agent [24]. However, the use of Iris is limited by the set of available commands and the need of an experienced user for solving data analysis problems. Third, the virtual assistant must support the collaborative process and make recommendations based on the analysis of the behavior of other users. There are many successful recommendation agent solutions, for example, in the field of e-commerce. However, for the CBI task they need to be adapted. Thus, the CBI virtual assistant must combine these three main CA directions.

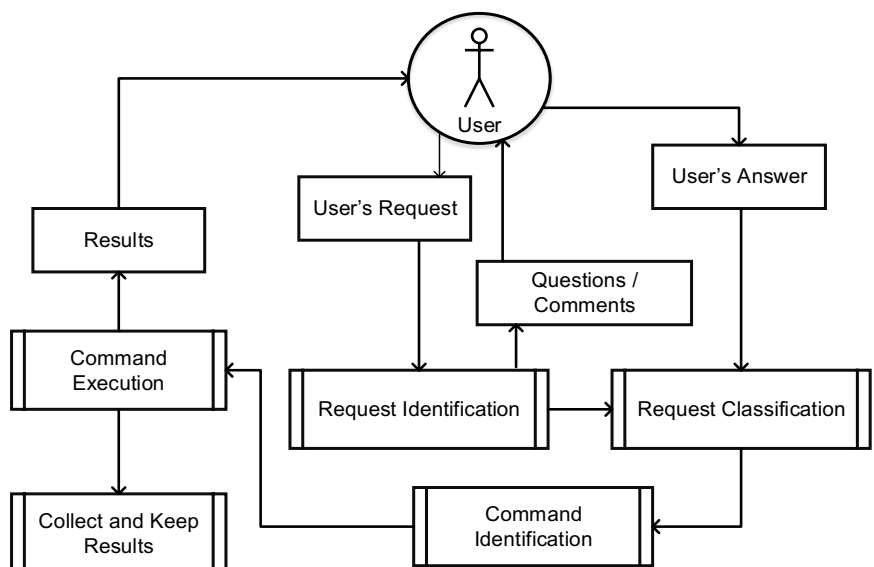

**Figure 2**: CBI dialogue example

The main idea of our research is to model CBI processes in distributed virtual teams via interaction of intelligent agents. In this paper, the agent is understood as a computer system placed in the external environment, able to interact with it, performing autonomous and rational actions to achieve a certain goal. The main difference between agents and systems in general is activity, i.e., the ability to perform any actions independently [46]. In addition, the agent is usually considered not as a set of parts, but as a single entity, while, for example, when studying the properties of systems, the first approach is the main one. Another characteristic is that the agent can be embodied not as a material object, but as a stand-alone program. However, this program, without affecting the material world (remaining within the computer or computer systems), can perform useful actions.

Fig. 3 depicts the CA, which consists of three components. They are the human-like interaction, data exploration and recommendation agents. The interaction agent is based on transfer learning. It communicates directly with the user and suggests some scenario of data research based on recommendations. The recommendation agent collects and processes a user's behavior and proposes to explore the data in order to achieve collaborative goals. The data exploration agent transfers expressions in natural language to the data processing commands.

Eventually, in order to construct a virtual assistant for CBI, it is essential to incorporate three main components: CA, data exploration and recommendation. It is recommended to develop these modules as a multi-agent system [46], which comprises three distinct types of agents that are suitable for their respective tasks. These agents provide support for dialogue with users, capture the context of data exploration and decision-making and offer recommendations. Our approach involves the interaction of the three types of agents, each of which is responsible for executing a specific task during the conversation.

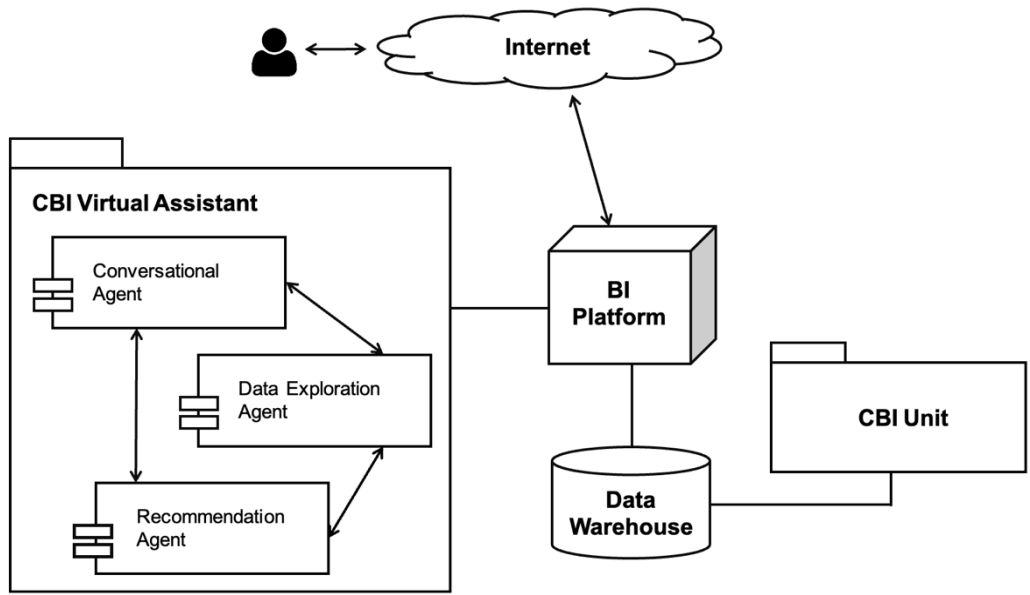

**Figure 3**: CBI virtual assistant reference model

## 6. Discussion

The state of the art of CBI can be described as a growing field with an increasing number of tools and techniques being developed to enable effective collaboration among teams in decision-making processes. CBI tools are becoming more user-friendly, accessible, and flexible, allowing users to customize their experience and adapt to their specific needs. Some of the most significant advancements in CBI include the integration of social media features, mobile accessibility, and cloud-based solutions. These developments have enabled users to work collaboratively and access data from any location, on any device, at any time. Additionally, the use of NLP and ML has made it easier for users to interact with data and extract insights, making decision-making processes more efficient and effective.

The state of the art in the domain of CA development has been rapidly evolving in recent years. The emergence of deep learning and NLP has allowed for the creation of more advanced and sophisticated CAs that can understand and respond to human language more accurately. Natural Language Querying (NLQ) can also make it easier for non-technical users to access and analyze data. This can help democratize data and insights across an organization, making them more accessible to a wider audience [4, 14, 47]. The goal of a virtual assistant is to make data exploration more accessible to a wider range of users and to reduce the time and effort required for data analysis. It can be used in various domains, including healthcare, finance, tourism, and e-commerce. It is an idea of creating innovative CAs is to transform the way users interact with data and ML models and to make data science more accessible to a wider range of users.

The Iris CA is based on a powerful combination of deep learning and reinforcement learning techniques, which allows to understand user queries and generate relevant responses [24]. This makes Iris a powerful tool for data scientists and other users who want to interact with data and ML models through natural language conversations. However, the limitation of Iris is a set of commands that it can transform and execute. It can help to data analytics who are not experienced a lot with Python libraries and Jupiter notebook, but such kind CA as Iris cannot to assist the collaborative decision-making.

Chatbots and CAs can be implemented as a part of self-service BI to enable users to easily access and analyze data without needing assistance from IT or data analysts. This allows users to ask questions in natural language and receive immediate responses, making data analysis more accessible and efficient. Our reference model is based on a combination of NLP and ML techniques, including deep learning and reinforcement learning as the cutting-edge tendency.

Thus, we can conclude that the virtual assistant can improve the collaborative tool by using deep reinforcement learning and NLP. The implementation of a pretrained model such as the Iris CA and fine turning towards data exploration and collaboration can provide the complete kit for powerful CBI.

## 7. Conclusion

CBI and CI are related concepts that both involve bringing people together to work collaboratively and share knowledge. CBI mostly refers to the use of collaborative tools and methods to support BI processes. This may include social networks, brainstorming sessions, chat applications and other technologies that facilitate knowledge sharing and collaboration between business users, analysts and technical specialists. The goal of CBI is to leverage the collective knowledge and insights of the group to improve decision-making, drive business outcomes and gain a competitive advantage. CI is a more general concept that applies to any situation where a group of people works together to solve a problem or make a decision. This may include scenarios outside of the business context, such as social or political movements, scientific research, or open-source software development. The goal of CI is to leverage the collective intelligence of the group to arrive at better solutions or make better decisions than would be possible with individual efforts alone.

Chatbots can be effectively used for CBI. They can be integrated with collaborative platforms or tools, such as chat applications or social networks, to facilitate knowledge sharing and collaboration between business users, analysts and technical specialists. There are many different applications where CAs can be beneficial. For example, CAs are being investigated to aid legal information access. Research is also being conducted regarding the role of CAs in vehicles and especially so in the realm of self-driving cars. Games are another application of relevance for CAs. Researchers have also investigated the effect of using CAs to help gaming communities grow and bond, by having the community members conversing with a CA in a platform. Finally, CAs are often proposed across different domains to act as recommender systems. CAs will likely become a technology that can offer many benefits. This technology may radically change the manner to interact with machines and shall allow all people the same access to online and digital services.

Our solution for CBI involves the use of virtual assistants. The goal of a virtual assistant is to make data exploration more accessible to a wider range of users and to reduce the time and effort required for data analysis. Towards this aim, we answer the research questions and can summarize the following.

1. The main scenario for a collaborative tool is virtual assistant based on text communicating involves presenting the findings in a clear and concise manner, using visualizations that can be easily understood by users. It includes dialogue between the user and chatbot in order to explore the data, choose the acts and create the visualizations. The chatbot has to provide recommendations and assist with data processing acts.

2. We choose three-component conversational agent that which consists of three units as conversational, data exploration and recommendation agents. They will interact each other and can be implemented as multi-agent module in order to support collaborative unit on the BI4People platform.

3. The AI models and natural language understanding should be identified in the future work based on experimental research due to wide set of available options.

## 8. Acknowledgments

The research study depicted in this paper is funded by the French National Research Agency (ANR), project ANR-19-CE23-0005 BI4people (Business intelligence for the people).